# Large Frame-Transfer Detectors for the MAIA Imager

Jesús Pérez Padilla[1,2], Simon Tulloch[3], Gert Raskin[2], Saskia Prins[1,2], Florian Merges[1,2], Wim Pessemier[2], Steven Bloemen[2,4]

[1]Mercator Telescope, Roque de Los Muchachos observatory, La Palma, Spain.
[2]Instituut voor Sterrenkunde, KU Leuven University, Leuven, Belgium.
[3]QUCAM, Cádiz, Spain.
[4]Department of Astrophysics, IMAPP, Radboud University Nijmegen, Netherlands.

**ABSTRACT**

MAIA, the Mercator Advanced Imager for Asteroseismology, is a new fast-cadence 3-channel photometric instrument. It is installed on the 1.2-m Mercator telescope at the Roque de Los Muchachos Observatory in La Palma, Spain.

MAIA comprises 3 cameras that simultaneously observe the same 9.4 x 14.1 arcmin field in 3 different colour bands (u, g and r). Each camera is based on a very large frame-transfer detector (CCD42-C0) of 2kx6k pixels, specially designed for rapid time-series photometry. These CCDs were originally developed by e2v for ESA's cancelled Eddington space mission and are now on permanent loan to the Institute of Astronomy of the KU Leuven, Belgium.

The acquisition system of MAIA uses a single ARC GEN-III controller, custom programmed to allow differing exposure times for each of the three CCDs. Predefined sequences synchronize each read-out with the shortest integration time. Detectors that are not read-out at the end of an exposure continue integrating and can be read-out in one of the subsequent cycles. This read method takes full advantage of the frame-transfer functionality of the CCD42-C0 detectors and allows optimisation of the exposure times for each wavelength band. This then gives a similar exposure depth in each of the 3 arms despite the fact that the UV channel typically sees much less flux.

We present the CCD42-C0 detectors, their characterisation, including a thorough analysis of their non-linearity, and the MAIA data-acquisition system.

## 1. INTRODUCTION

The MAIA project found its origin in the Eddington space mission that was cancelled by the European Space Agency (ESA) in 2003. A number of charge-coupled device (CCD) detectors had been developed and procured specifically for this mission, based on the well-known e2v CCD42-xx series. These are large frame-transfer devices, designated as CCD42-C0 [1].

After the cancellation of the Eddington mission, the Institute of Astronomy of the University of Leuven (KU Leuven, Belgium) received the Eddington detectors through a permanent loan agreement between ESA and KULeuven.

The present paper presents the design and configuration of the acquisition system for the MAIA imager that is based on three of these large frame-transfer detectors. We also discuss the custom programmed functions of the CCD controller that are needed to operate these detectors in a fast-cadence mode as required for this instrument (MAIA). Finally, we present the detector

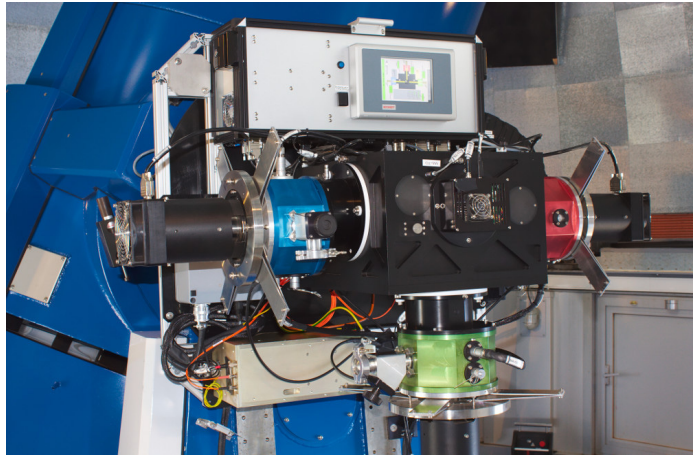
*Figure 1 MAIA instrument at the Nasmyth focus of the 1.2-m Mercator telescope.*

characterisation and the optimisation of the linearity, improving the behaviour at low levels of signal.

## 2. THE MAIA INSTRUMENT

### 2.1 Science case and requirements for acquisition system

With asteroseismology one can constrain the internal structure of stars by means of a frequency analysis and mode identification of the excited harmonics of pulsating stars [2]. The observational basis is most often an accurate and extensive photometric time-series.

The amplitude of pulsation modes in different colours depends on the spherical degree of the modes. In order to measure the ratios of pulsation amplitudes in different colours to the precision necessary to identify the spherical degree of the pulsation modes, the photometric variations must be measured simultaneously in multiple bands and with respect to constant reference stars in the field. This requires instrumentation capable of simultaneous multi-band imaging over a large field of view and an acquisition system that is optimised for short cycle times.

Many possible MAIA targets are faint or span a broad brightness range. This is especially relevant when considering the different wavelength ranges that are sampled simultaneously. For a typical reference star, the *u*-band flux can be very small and many magnitudes fainter than the *r* or *g*-band flux. Therefore the acquisition system must provide a synchronized method to allow different exposure times for each of the three detectors.

### 2.2 Instrument layout

MAIA, the Mercator Advanced Imager for Asteroseismology, is a new fast-cadence photometric instrument that is installed on the 1.2-m Mercator telescope at the Roque de los Muchachos Observatory in La Palma, Spain [3]. MAIA is a three-channel imager that observes simultaneously the same 9.4 x 14.1 arcmin$^2$ field of view in three different colour bands (*u*, *g* and *r*). MAIA consists of

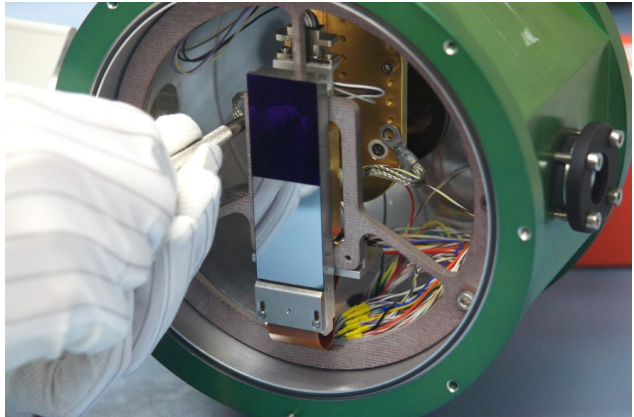
*Figure 2  Picture of a CCD42-C0 detector in the g cryostat.*

a common collimator, two dichroic beam splitters, three science cameras and an additional beam splitter to serve the guiding camera. In 2012, MAIA was installed on the 1.2-m Mercator telescope at the Roque de los Muchachos Observatory on La Palma (Canary Islands, Spain), and commissioning took place in 2013.

## 3. DETECTORS

Each of the three cameras of MAIA is based on a frame-transfer (FT) detector (CCD42-C0), originally developed by e2v for ESA's cancelled Eddington space mission. One of the characteristics of these devices is their large format of 2048 x 6144 13.5µm pixels, split in halves between a 2k x 3k imaging area, and an equally large storage area. Figure 2 shows a picture of a CCD42-C0 detector during its assembly in the *g* cryostat, where we can see the photo-sensitive imaging area (dark), and the storage area (aluminium-covered). Detector pin-out is very similar to the known CCD42 series, except for the three frame-transfer clock lines. These parallel clock lines allow a shift of the charge from the photo-sensitive image area to the storage area in less than 300 ms, turning this detector in an excellent device for rapid time-series photometry.

## 4. ACQUISITION SYSTEM

The detector system of MAIA relies on a single ARC GEN-III SDSU controller (Astronomical Research Cameras, USA), equipped with two double-channel ARC-45 Video boards, two ARC-32 Clock Driver boards and one ARC-22 Timing board. To save on the number of boards in the detector controller, only single-port readout is allowed for each of the three detectors. The r and g cameras use only one channel of the first video board and the u camera is connected to the first channel of the second video board. Each cryostat is equipped with two CCD connectors, one for left-port and the other for right-port read out. The three detectors also share the clock driver boards. One of them is used for frame transfer and parallel clocks, which can operate independently for each of the three detectors. The second clock driver board controls the serial registers lines to read the storage area of the three detectors simultaneously. Cross-talk and noise were checked in this shared configuration and no significant effect was observed.

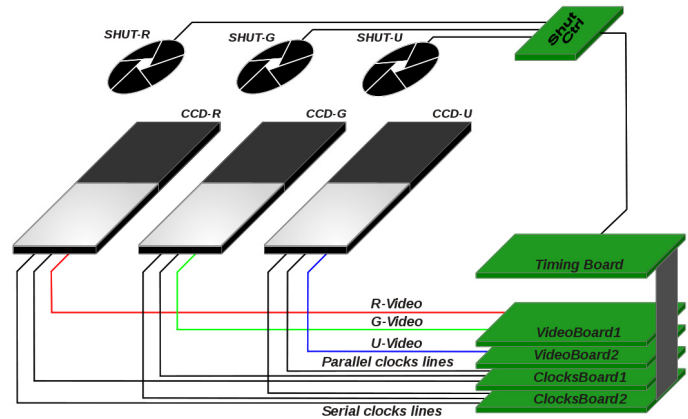

*Figure 3 Overview of the science detector system.*

### 4.1 System characteristics

The system was specifically designed to allow the use of the frame-transfer characteristic for the three detectors independently, including a fast shift of the image data from the imaging to the storage area. With this utility a new integration starts in the photo-sensitive area just after the shift, while the previous image is being read and digitised from the storage area by the controller. The programmed frame transfer shift takes only 295 ms, while the readout of a full frame takes more than 43 seconds in the default speed mode (LGN, see table 1). These numbers show that e.g. for sequential exposures with 45 seconds integration time, the acquisition system has a duty cycle of 99.3%, as the dead time due to the FT shift is only 295 ms.

The usage of the FT in a sequential mode implies that the integration time should be at least as long as the total duration of the image digitisation. For shorter integration times, it is possible to disable the FT option and clear the detectors at the beginning of each new exposure.

*Table 1 Read-out mode characteristics.*

|  | Slow (LGN) | Fast (MGN) |
|---|---|---|
| **Read-out frequency** | 156 kpixel s$^{-1}$ | 227 kpixel s$^{-1}$ |
| **Read-out noise** | 3.5 – 4 e$^{-}$ | 4.5 – 5 e$^{-}$ |
| **Single row read-out time** | 13.7 ms | 9.5 ms |
| **Full-frame read-out time** | 43.1 s | 29.8 s |
| **Full frame transfer time** | 295 ms | 295 ms |
| **Conversion gain** | 0.8 ADU/e$^{-}$ | 0.4 ADU/e$^{-}$ |

As explained earlier, during sequential exposures in FT mode, a shutter is not needed to obscure the detector during read-out. Nevertheless, we decided to equip the three MAIA cameras with a mechanical iris shutter.

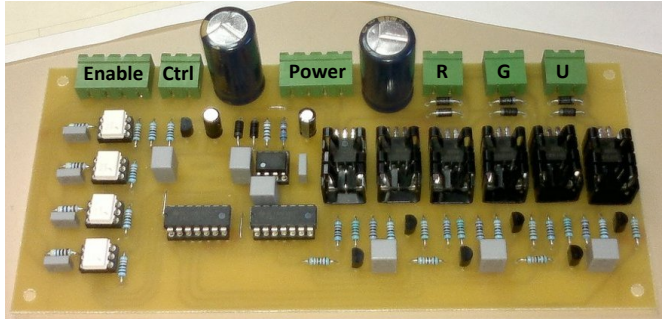

*Figure 4  Shutter control board.*

During FT observations, the shutter is continuously open until the end of the last exposure in the sequence. The shutter control board also has the option of disabling any of the three shutters individually, which is useful to protect the detectors from over-illumination in specific conditions, e.g. closing the *g* and *r* shutters to avoid strong over-exposure during *u* sky flat fields.

## 4.2  Custom programmed functions for MAIA

### 4.2.1  Exposure sequences

The data-acquisition system has been designed and custom programmed to allow differing exposure times for each of the three CCDs. In the Timing board, predefined sequences synchronize each read-out with the shortest integration time, taking full advantage of the frame-transfer functionality. Detectors that are not read-out at the end of an exposure continue integrating and can be read-out in one of the subsequent cycles. This allows the optimisation of the exposure times for each wavelength band, since the user can select a multiple (x1, x2 or x4) of the nominal exposure time, and then the controller will generate an internal sequence of simultaneous exposures for the three channels. This characteristic helps to give a similar exposure depth in each of the 3 arms, despite the fact that the *u* channel typically sees much less flux. As an example, a sequential exposure with a base time of 10 s, and a programmed sequence of x1, x2, x4 for the channels *r*, *g* and *u*, respectively, will generate four 10s exposures through the *r*-channel, two 20s *g*-channel exposures and one 40s *u*-channel exposure.

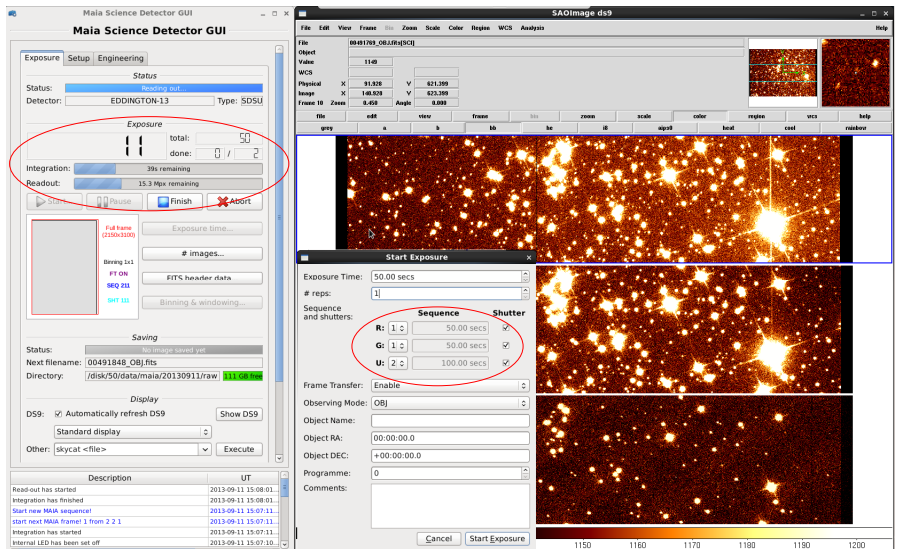

*Figure 5  Graphical user interface for frame-transfer exposures.*

Figure 5 shows the Graphical User Interface with the controller running in frame-transfer mode. Integration time and readout progress bars indicate the remaining time during exposures. The *Start Exposure* dialog box serves to configure, prior to the first integration, the exposure base time, the different exposure times for each of the three channels, the number of repetitions for the sequential loop, the option to disable or enable the shutters, etc.

### 4.2.2 Multi-Windowing and binning

Standard single windowing of the detectors as well as a multi-window mode are implemented. Both windowing modes make use of fast parallel-skip clocks to dump the charge below the windows and clear the serial register just before dumping the first row of each window. In the actual controller code the number of windows in multi-windowing mode is limited to ten, and all of them must have the same dimensions.

For the moment, horizontal windowing is not allowed because we found that serial skips cause a disturbing bias level gradient at the start of each window. Therefore, horizontal extent of the windows has been fixed to the full width of the detector, including prescan and overscan (2150 pixels). This issue will be addressed in the future for the implementation of horizontal windowing.

When combining the frame-transfer mode with windowing of the detector, the single-row read-out time can be used to determine the read-out and minimum exposure times by multiplying with the number of window rows (table 1).

### 5. LINEARITY STUDY AND OPTIMISATION

During the commissioning of the instrument, we found that the detectors showed non-linear behaviour. The photometric response was not linear below 300 ADU ($\approx$400 e$^-$) for the *g* channel, and below 100 – 160 ADU ($\approx$125 – 200 e$^-$) for the *r* and *u* channels. Solving this problem was really imperative for the *u* detector due to the low signal levels that we expect in this channel. Figure 6 shows the response without optimisation.

To characterize this non-linearity, a photometric study was carried out. Gain, gain error and flux were carefully measured in extended series of images with a broad range of exposure times. These images were obtained by illuminating the detector with an internal LED in each cryostat to avoid filter and shutter effects, as well as from standard dome flat-fields images through the telescope and instrument optics. As the LED exposures have a strong intensity gradient, each image offers a broad range of signal levels. This makes them very well suited for this type of analysis.

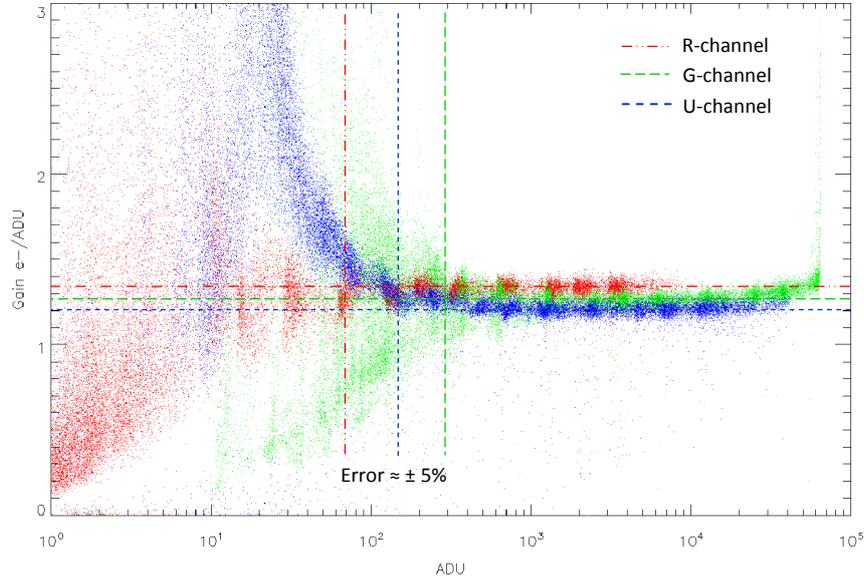
*Figure 6 Gain versus signal: without optimization, the response is non-linear.*

To accurately assess detector linearity at these low signal levels, the data-reduction requires great care to remove the effects of the zero-second pixel background and the effects of gradients in the bias signal. The correct signal level *S* in the window of interest is determined using the following equation:

$$S\ [ADU] = \text{mean(image Window)} - \frac{1}{2}\left(\text{mean(prescan)} + \text{mean(overscan)}\right) - \left(\text{mean(biasframe Window)} - \frac{1}{2}\left(\text{mean(prescan)} + \text{mean(overscan)}\right)\right)$$

The gain values *G* are obtained from two images, taken with exactly the same exposure time and with read noise *RN*:

$$G\ [ADU/e^-] = \frac{\left(\text{StdDev}(\text{Image1} - \text{Image2})\right)^2 - 2 \cdot RN^2}{2 \cdot S}$$

Figure 6 shows the non-linear response and variable gain at low signal levels that we found using this analysis method.

## 5.1 Hardware level improvements

The first step towards improving the linearity of the system was the optimisation of the bias voltages that are applied to the output transistors of the CCDs. In the case of the OD (output drain) pins of the CCDs, current was supplied through 1kΩ series resistors on the video boards, across which at least one volt was lost. Reducing these resistors to 100 Ω produced a small improvement in the behaviour of the system but a substantial amount of non-linearity remained.

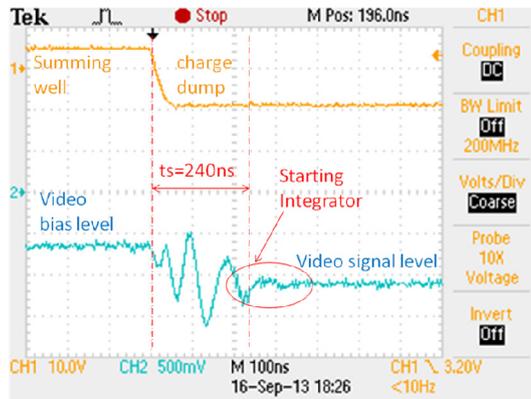

*Figure 7   Summing well clock (CH1) and video waveform (CH2) of the g channel (JFET output) before optimization.*

### 5.2   Clocks and video waveforms study

We also checked if the non-linearity was related to the parallel and serial clock frequencies. The waveform frequencies are set in accordance to e2v's *Clock waveforms technical note* for the CCD42-C0 and slower clocking rates do not improve the linearity behaviour.

Subsequently, an oscilloscope investigation of the video signals revealed much longer signal settling times than expected. In the original version of the timing board code, the controller started the integrator 240 ns after "charge dump". Figure 7 shows a scope capture of the video waveform for the *g*-channel. Settling time is clearly longer than the expected 240 ns.

In the same technical note about the CCD42-C0, we find the reference values for the video output settling time (*ts*) as a function of the load capacitance (Table 2). The reason for this longer settling time may be due to a higher capacitance in the video board input, a higher serial inductance on the cables, or a combination of both.

*Table 2   CCD42-C0 Output parameters.*

| Condition | $t$ = 10-90% | $ts$(1%) |
|---|---|---|
| Device loaded with JFET | 45 ns | 90 ns |
| Direct output, 12pF load | 55 ns | 110 ns |
| Direct output, 22pF load | 70 ns | 140 ns |
| Direct output, 59 pF load | 105 ns | 210 ns |
| Direct output, 112pF load | 165 ns | 330 ns |

In addition, these clock transitions always cause a significant feed-through to the output video signal, independently of the photo-electric signal. When sampled by the video processor, this will add non-linearity to the video amplifier response.

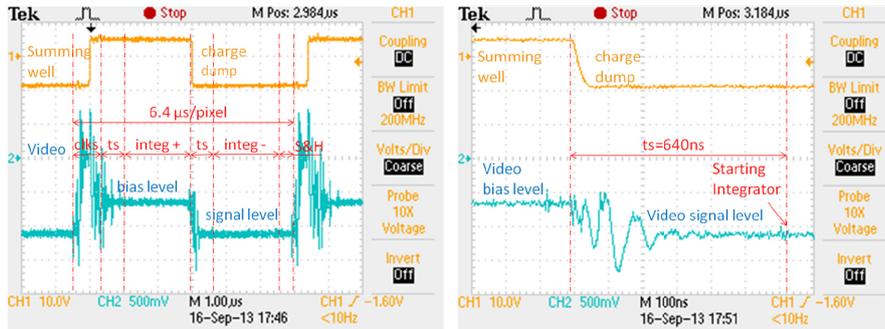

*Figure 8  Left: summing well clock (CH1) and Video waveform (CH2) with the optimized timing code; right: detailed view of video waveforms after charge dump with ts = 640ns.*

## 5.3  Pixel digitisation waveforms

The timing board code and the clock waveforms timing were iteratively improved, increasing the delay before starting the bias level integration (+) and before the signal level integration (–), in 120 ns steps. Finally, with a delay of ts=640 ns before starting each integrator ramp, we obtained a system with satisfying linearity behaviour for all three channels. The video waveforms produced by the final timing code are shown in Figure 8.

The linearity results obtained with the final version of the timing code are excellent, as indicated by the photometric analysis shown in Figure 9. A series of flat-field images was taken with exposure times ranging from 30 ms to 10 s in a geometrically spaced sequence. The response is linear down to signal levels of 10 e⁻.

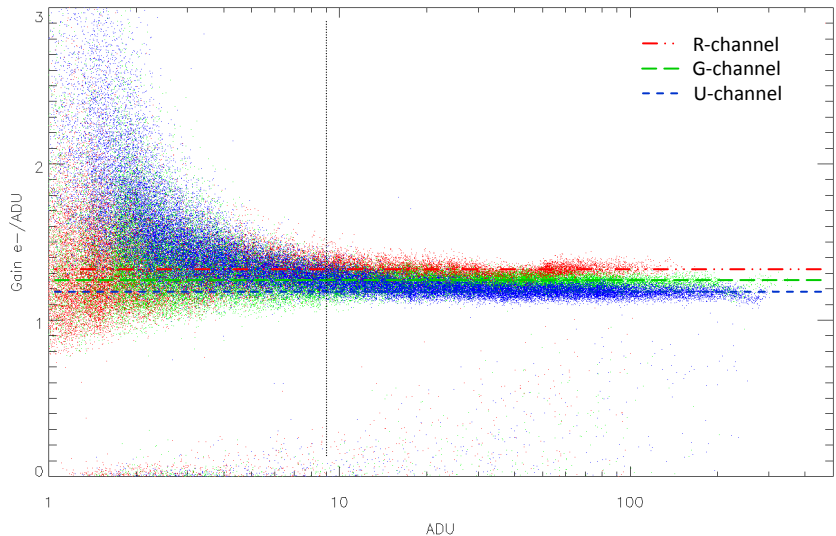

*Figure 9  Gain versus signal: after optimization, the response is linear down to signal levels below 10 ADU.*

## 6. CONCLUSIONS

With the construction of the MAIA imager for the Mercator telescope, the unique frame-transfer CCDs that were developed for the Eddington space mission are finally being used for one of their original science goals, albeit from a ground-based telescope. We have developed a data-acquisition system for these detectors that exploits them in an efficient way for fast-cadence time-series photometry.

Initially, the Eddington detectors showed a strong non-linear response that compromised their usability for precision photometry. These linearity problems were solved through small hardware adjustments and an optimisation of the controller timing code, both aided by an accurate photometric analysis method. The result is a detector system with excellent linearity down to very low signal levels.

## 7. ACKNOWLEDGEMENTS

This research was based on funding from the European Research Council under the European Community's Seventh Framework Programme (FP7/2007–2013)/ERC grant agreement n°227224 (PROSPERITY) and from the Fund for Scientific Research of Flanders (FWO), grant agreements G.0410.09 and G.0470.07, and the Big Science program. The CCDs of the MAIA camera were developed by e2v in the framework of the Eddington space mission project and are owned by the European Space Agency; they were offered on permanent loan to the Institute of Astronomy of KU Leuven, Belgium, with the aim to build and exploit an instrument for asteroseismology research to be installed at the 1.2m Mercator telescope at La Palma Observatory, Canary Islands.

## 8. REFERENCES

[1] G. Raskin, S. Bloemen, J. Morren, et al. "MAIA, a three-channel imager for asteroseismology: instrument design", Astronomy & Astrophysics, Vol. 559, pp. A26, 2013.

[2] Aerts, C., Christensen-Dalsgaard, J., and Kurtz, D. W., 2010, Asteroseismology, Springer Netherlands.

[3] G. Raskin, G. Burki, M. Burnet, et al. "Mercator and the P7-photometer", Society of Photo-Optical Instrumentation Engineers (SPIE) Conference Series, Vol. **5492**, pp. 830, 2004.